\documentclass[11pt]{article}
\usepackage{jheppub,amsmath, amsthm, amssymb,slashed,url}
\usepackage{amsfonts,amsthm,bm,booktabs,mathtools}
\usepackage{graphicx}
\usepackage{epstopdf}
\usepackage{tikz}
\usepackage{xcolor}
\usepackage{tabu}
\usepackage{tabularx}

\usepackage{etoolbox}
    \makeatletter
    \patchcmd{\maketitle}{\@fpheader}{}{}{}
    \makeatother

\usepackage{etoolbox}
\notoc

\makeatletter

\patchcmd{\maketitle}{\@fpheader}{}{}{}

\patchcmd{\maketitle}
  {\ifnotoc\else}
  {\ifnotoc\newpage\else}
  {}
  {\PackageError{titlefix}{Could not patch maketitle}{}}

\makeatother

\usepackage{boldline}
\usepackage{bm}
\usepackage{amssymb,amsfonts}

\newcommand{\Tr}{\operatorname{Tr}}

\newcommand{\NL}{\mathrm{NL}}

\newtheorem{theorem}{Theorem}[section]

\newtheorem{lemma}[theorem]{Lemma}

\def\OMIT#1{{}}

\newcommand{\beq}{\begin{equation}}
\newcommand{\eeq}{\end{equation}}
\newcommand{\beqa}{\begin{eqnarray}}
\newcommand{\eeqa}{\end{eqnarray}}

\dedicated{IQuS@UW-21-135}
\title{Local universality of non-local quantum resources  in rotationally invariant scattering}
\author[a]{\bf Silas R.~Beane}
\author[b]{\bf and Roland C.~Farrell}

\affiliation[a]{InQubator for Quantum Simulation (IQuS) and Department of Physics,\\
  University of Washington, Seattle, WA 98195.}
\emailAdd{silas@uw.edu}
\affiliation[b]{Institute for Quantum Information and Matter (IQIM) and Department of Physics,\\
California Institute of Technology, CA 91125, USA}
\emailAdd{rolandf@caltech.edu}
\date{\mydate}

\abstract{
Elastic scattering can generate entanglement and non-local magic between the spins of the two outgoing particles.
The amount of entanglement and magic produced is encoded in the S-matrix; a unitary quantum gate that acts on the joint Hilbert space of the two spin-$J$ particles.
Previous work derived the entanglement power in terms of the scattering phase shifts for rotationally invariant S-matrices with $J=1/2,1,3/2$.
In this paper, a closed form solution is derived for all $J$.
We further prove that a broad class of non-local quantum resources depend to leading order only on the squared projective Hilbert–Schmidt distance from the nearest non-entangling gate. 
This holds for the entanglement power and non-local magic power, and the choice of resource simply changes the overall normalization. This universality establishes a common geometric origin for the leading resource-generating capacity of scattering, independent of the details of the interaction.
}

\begin{document} \maketitle

\section{Introduction}
\label{sec:introduction}
\setcounter{page}{1}
The elastic scattering matrix (S-matrix) is a unitary quantum gate
whose capacity to generate quantum resources is related to
the symmetries of the underlying interaction. This observation was
made originally in the context of low-energy baryon-baryon scattering
where it was shown that the entanglement power of the S-matrix in spin
space vanishes at points of enhanced global symmetry that are approximately realized in nuclear
data~\cite{Beane:2018oxh}.  
The fact that nature is near
entanglement-free regions of the resource landscape suggests that extremization
of a quantum resource is a deeper principle from which ``accidental" symmetries could be
derived~\cite{Beane:2018oxh,Low:2021ufv,Beane:2021zvo,Liu:2022grf,Carena:2023vjc,Hu:2025lua,Miller:2023ujx,Kanemura:2026won}. 
The entanglement power as a function of the scattering phase shifts was first derived for two distinguishable particles with spin (or isospin) $J=1/2$~\cite{Beane:2018oxh} and later extended to $J=1$~\cite{Beane:2021zvo,Kirchner:2023dvg} and $J=3/2$~\cite{Hu:2025lua}.
Very recently~\cite{Low:2026mxh}, the entanglement power of an arbitrary unitary was computed at stationary points where all relative eigenphases are 0 or $\pi$, corresponding to S-matrices where all phase shift differences are $\delta_\mu -\delta_\nu \in \{0,\pi/2\}$.
In this paper, we give the general solution for the entanglement power that is valid for any $J$ and all phase shifts.
This specializes the entanglement power of an arbitrary unitary~\cite{PhysRevA.63.040304,Zanardi:2000zz} to a rotationally invariant S-matrix.

The resource landscape of scattering can be extended to include non-stabilizerness or magic, a distinct measure from entanglement that is particularly relevant to the complexity of fault tolerant quantum computation~\cite{Bravyi:2004isx,Gottesman:1998hu}.
The magic power
of the two-baryon S-matrix was computed in Ref.~\cite{Robin:2024oqc}
and that analysis was extended to include non-local magic and
anti-flatness of the entanglement spectrum ---quantities which survive
a local change of basis--- in Ref.~\cite{Robin:2026lnm}, see also Ref.~\cite{Qian:2025oit}. A generalization of the resource
landscape in the nucleon-nucleon system to higher partial waves appears in
Ref.~\cite{Low:2026evp}. For a recent review of quantum complexity
with applications in nuclear and particle physics, see
Ref.~\cite{Robin:2026lqp}.

The presence of distinct quantum resources
raises the question: in what sense are they independent diagnostics of
quantum complexity? 
We answer this question in the vicinity of S-matrices that have zero entanglement power.
Non-entangling two-qudit unitaries are local unitaries or local unitaries composed with SWAP~\cite{Johnston:2011svd}, a classification applied to scattering in Ref.~\cite{McGinnis:2025brt}. 
For the rotationally invariant spin S-matrices considered here, the class is restricted to identity and SWAP, up to an overall phase.
In this region, we prove that a wide class of non-local resource functions, including the entanglement and non-local magic power, are equivalent up to overall normalization. 
Therefore, the leading non-local resource-generating capacity of a rotationally invariant S-matrix is universal.
Related connections between non-local magic and entanglement have been established in Refs.~\cite{Andreadakis:2025mfw,Torre:2026pil,Huang:2026mhv} for qubit systems.

We additionally show that the leading scaling of non-local resources is fundamentally tied to the geometry of the space of all S-matrices~\cite{Beane:2020wjl,Beane:2021bpr}; it measures the squared projective Hilbert-Schmidt distance from the nearest non-entangling gate.
This is a much stronger constraint than comes from expanding a resource function around zero, as a resource can generically grow at different rates along different directions. 
Our proof shows that the non-local resource Hessian matrix is proportional to the Hilbert–Schmidt metric and, consequently has no directional dependence.  

This paper is organized as follows.
In Section~\ref{sec:Smat_EP}, the S-matrix is introduced and the closed form expression for the spin-$J$ entanglement power is given.
Details on the derivation are provided in Appendix~\ref{app:EP_SVD}.
Section~\ref{eq:Identity_Expand} shows that the entanglement and non-local magic power near the identity and SWAP are proportional, with details given in Appendix~\ref{app:EP_Magic_Identity}.
In Section~\ref{sec:HS_Metric}, the projective Hilbert Schmidt distance on the space of S-matrices is introduced and shown to measure the leading non-local resource generation in the vicinity of identity and SWAP. The proof is given in Appendix~\ref{app:resource_proof}.
Section~\ref{sec:discussion} discusses the results.

\section{The entanglement power for rotationally invariant scattering}
\label{sec:Smat_EP}
Consider elastic
scattering between two distinguishable non-relativistic particles of spin $J$.  
The interaction is assumed to be
rotationally invariant and of finite range. Let $\mathcal H_J$
be the spin-$J$ irreducible Hilbert space. 
The two particles span the space $\mathcal H_J\otimes\mathcal H_J$, with local qudit dimension
$d=2J+1$. 
Angular-momentum addition decomposes this space
into sectors of total spin, and the elastic s-wave S-matrix takes the form
\begin{equation}
 S(\bm\delta)=\sum_{\mu=0}^{2J}e^{2i\delta_\mu}P_\mu \ ,
 \qquad \Tr P_\mu=r_\mu=2\mu+1 \ ,
 \label{eq:Smatgate}
\end{equation}
where $\mu$ is total spin, $P_\mu$ is the orthogonal projector onto the spin-$\mu$ sector,
$r_\mu=2\mu+1$ is its rank, and the real, dimensionless phase shift
$\delta_\mu$ is defined modulo $\pi$. 
We have assumed that the s-wave does not couple to higher partial waves so that the S-matrix is unitary.
The projectors are fixed by representation theory; all
dynamical information is contained in the phase shifts.

In elastic scattering with finite-range interactions, particles approach from asymptotic infinity, interact, and return to asymptotic infinity. 
We assume that the initial spin states, $|\psi_A\rangle$ and $|\psi_B\rangle$, are prepared in independent pure states.
The S-matrix connects the incoming state (in) to the outgoing state (out),
\begin{align}
 |\text{in}\rangle\ = \ |\psi_A\rangle\otimes|\psi_B\rangle \ , \quad |\text{out}\rangle \ = \ S|\text{in}\rangle 
\  .   
\end{align}
To characterize the resource-generating capacity of the collision, each resource is averaged over independently sampled pure states of $|\psi_A\rangle$ and $|\psi_B\rangle$.
The state of each spin is a normalized complex $d$-dimensional vector, with vectors differing only by an overall phase identified. 
The state manifold is the complex projective space $\mathbb{CP}^{d-1}$ and each incoming state is sampled from a normalized distribution induced by the Haar measure over unitary matrices. 
This provides a common initial state ensemble for the entanglement and magic measures used in this work.
It differs from previous studies of magic that sampled initial states from a discrete stabilizer ensemble~\cite{Robin:2024oqc,Robin:2026lnm,Low:2026evp}. 

The
entanglement power~\cite{Zanardi:2000zz} is the average linear entropy of the out state,
\begin{equation}
 \mathcal E_{\rm p}(S)\ = \ \int d\psi_A\,d\psi_B\,
 \left[1-\Tr\rho_A^2\right] \ ,
 \qquad
 \rho_A\ = \ \Tr_B\!\left(|\text{out}\rangle\langle\text{out}|\right) \ .
 \label{eq:EPdef}
\end{equation}
The 
$d\psi_A$ and $d\psi_B$ are normalized Haar probability measures on $\mathbb{CP}^{d-1}$ and $\rho_A$ is the reduced density matrix on subsystem $A$.

Before stating the final result, some intuition is gained by performing a particular Schmidt decomposition of the S-matrix.
First, write the S-matrix as a four index tensor that acts separately on particle A and particle B,
\begin{align}
    S \ = \ \sum_{ijab}S_{ij,ab}|i\rangle \langle a|\otimes |j\rangle\langle b| \ .
\end{align}
Indices $(a,b)$ are associated with the incoming particles, and $(i,j)$ with the outgoing particles.
For $J=1/2$, a familiar choice of basis would be the tensor product of orthonormal Pauli operators $\frac{1}{2}\{I,X,Y,Z\}_{ia}\otimes\{I,X,Y,Z\}_{jb}$.
The coefficients $S_{ij,ab}$ can be represented by a $d^2\times d^2$ matrix $S_{I,A}$ using the standard index mapping $I =di + j$ and $A=da+b$.
The coefficient matrix can be regrouped into particle A $(i,a)$, and B $(j,b)$,
\begin{align}
    S \ = \ 
    \sum_{iajb}{\cal R}(S)_{ia,jb}|i\rangle\langle a|\otimes |j\rangle\langle b| \ . 
\end{align}
This is a realignment with the indices of ${\cal R}(S)_{I,J}$ given by  $I =di + a$ and $J=dj+b$.
The row (column) of ${\cal R}(S)$ is associated with particle A (B) and its singular values are preserved under unitary transformations that act separately on the left (particle A) or right (B).
An S-matrix whose ${\cal R}(S)$ only has one nonzero singular value is non-entangling as its action factorizes between the two particles.
Therefore, intuitively, the entanglement power should be sensitive to the number of non-zero singular values as well as their magnitude.

The entanglement power of an arbitrary unitary acting on a bipartite qudit Hilbert space was given in Ref.~\cite{PhysRevA.63.040304}. Applying this general formula to the S-matrix gives,
\begin{align}
    \mathcal E_{\rm p}(S) \ = \ \frac{d^2+1}{(d+1)^2} \ - \ \frac{{\cal C}(S) \ + \ {\cal C}(S\,\text{SWAP})}{d^2(d+1)^2} \ ,
    \label{eq:EP_general}
\end{align}
where
\begin{align}
    {\cal C}(U) \ = \ \Tr{\left [({\cal R}(U){\cal R}(U)^{\dagger})^2 \right ]} \ = \ \sum_{\alpha} |\gamma_{\alpha}|^4 \ ,
\end{align}
with $\gamma_\alpha$ the singular values of ${\cal R}(U)$.
Thus, the dynamical content of the entanglement power comes from the singular values of the realigned matrices ${\cal R}(S)$ and ${\cal R}(S\,\text{SWAP})$ in the basis that splits particle A and B~\cite{Ma:2007tny,Lu:2008apn}. 
The definition in Eq.~\eqref{eq:EPdef} makes it clear that the entanglement power is invariant under $S\to S\,\text{SWAP}$ as this just relabels the integration measures $d\psi_A \leftrightarrow d\psi_B$.
This symmetry is manifest in Eq.~\eqref{eq:EP_general} and explains the presence of both ${\cal C}(S)$ and ${\cal C}(S\,\text{SWAP})$.

A rotationally invariant S-matrix is block diagonal in total spin. This extra structure allows ${\mathcal C}(S)$ and ${\mathcal C}(S\,\text{SWAP})$ to be computed in closed form.
In Appendix~\ref{app:EP_SVD} it is shown that,
\begin{align}
    {\cal C}(S) \ = \ \sum_{L=0}^{2J} (2L+1)|a_L|^4\ , \qquad  {\cal C}(S \,\text{SWAP}) \ = \ \sum_{L=0}^{2J} (2L+1)|b_L|^4 \ ,
    \label{eq:CsCsSwap}
\end{align}
where 
\begin{align}
    a_L \ = \ \sum_{\mu=0}^{2J} (2\mu+1)(-1)^{2J-\mu}  e^{2i\delta_{\mu}} \begin{Bmatrix} J & J & \mu \\ J & J & L \end{Bmatrix} \ , \qquad b_L \ = \ \sum_{\mu=0}^{2J} (2\mu+1) e^{2i\delta_{\mu}}\begin{Bmatrix} J & J & \mu \\ J & J & L \end{Bmatrix} \ . 
\end{align}
The $\{.\}$ symbol is the Wigner $6j$ symbol
and
is given explicitly in terms of a finite sum by the Racah formula~\cite{NIST:DLMF}.
The Wigner $6j$ symbol has previously been used to connect the total-spin projector basis to a basis built from products of local operators in Ref.~\cite{Breuer:2005vwj}.
Putting these terms together, the entanglement power for arbitrary spin-$J$ is,
\begin{align}
    \mathcal E_{\rm p}(S) \ = \ \frac{d^2+1}{(d+1)^2} \ - \ \frac{\sum_{L=0}^{2J}(2L+1)(|a_L|^4 + |b_L|^4)}{d^2(d+1)^2} \ .
    \label{eq:EP_anyJ}
\end{align}

\section{Entanglement and magic near the identity and SWAP}
\label{eq:Identity_Expand}
The identity and SWAP are the only two S-matrices with zero entanglement power, up to an overall phase~\cite{McGinnis:2025brt}. 
To describe a small perturbation from $S_0=I$ or $S_0=\text{SWAP}$, write
\begin{align}
    S(\tau)=S_0e^{i\tau K} \ ,
\qquad
K=\sum_\mu\epsilon_\mu P_\mu \ ,
\end{align}
so that channel $\mu$ acquires an additional phase $ \tau\epsilon_\mu$. Removing the common phase gives
\begin{align}
    \widetilde K=K-\bar\epsilon I \ ,
\qquad
\bar\epsilon=\frac{1}{d^2}\sum_\mu r_\mu\epsilon_\mu \ , \qquad \Tr{\widetilde K}=0 \ .
\end{align}
With this parameterization it is straightforward to evaluate the singular values and derive the entanglement power to lowest order in $\tau$.
The result is,
\begin{align}
    \mathcal E_{\rm p}(S) \ = \ \frac{2\tau^2}{(d+1)^2}\Tr{\widetilde K^2} \ + \ O(\tau^4)
    \label{eq:EP_LO}
\end{align}
with details given in Appendix~\ref{app:EP_Magic_Identity}. 
The dependence on the scattering phase shifts comes from $\Tr{\widetilde{K}^2}$.

Exact expressions for the non-local magic can also be derived in the vicinity of the identity and SWAP.
Let $D=d^2$ be the dimension of the two-spin Hilbert space. 
We denote by
\begin{align}
    \mathcal{P}_D \ = \ \{P_A\otimes P_B : P_A\in\mathcal{P}_A,\ P_B\in\mathcal{P}_B\}
\end{align}
the set of $D^2$ chosen Pauli--Weyl representatives, one for each element of the Pauli--Weyl group up to a scalar phase.
Our conventions are given in Appendix~\ref{app:EP_Magic_Identity}.
A pure stabilizer state is the unique common (+1) eigenstate of a commuting subgroup of the full Pauli group, including its scalar phases. 
The linear stabilizer magic of a state is~\cite{Leone:2021rzd}
\begin{equation}
 M_{\rm lin}(|\psi\rangle)
 \ = \ 1-F_{\rm stab}(|\psi\rangle) \ ,
 \qquad
 F_{\rm stab}(|\psi\rangle)
 \ = \ \frac1D\sum_{P\in\mathcal P_D}|\langle\psi|P|\psi\rangle|^4 \ . 
 \label{eq:linearMagic}
\end{equation}
With this normalization,
$M_{\rm lin}=0$ on every stabilizer state.  
The non-local linear magic is defined by
\begin{equation}
 M_{\NL}(|\psi\rangle)
 \ = \ \min_{U_A,U_B}M_{\rm lin}\big((U_A\otimes U_B)|\psi\rangle\big) \ ,
 \label{eq:NLMdef}
\end{equation}
where $U_A,U_B\in U(d)$ are arbitrary local unitaries acting on the Hilbert space of $A$ and $B$ respectively.
The
associated gate power averages this quantity over the ensemble of initial product states
\begin{align}
    \mathcal P_{\NL}(S)\ = \  \int d\psi_A\, d\psi_B
M_{\rm NL}\left(|\text{out}\rangle\right)\ .
\end{align}

An S-matrix near the identity or SWAP leaves the outgoing state close to a product state. More precisely, the largest weight in the Schmidt decomposition is close to 1, i.e.
\[
|\text{out}\rangle \ = \ \sum_i\sqrt{\lambda_i}|u_i\rangle\otimes|v_i\rangle \ , \qquad \lambda_0=1\ - \ \eta 
\]
with $\eta$ a small parameter.
In Appendix~\ref{app:EP_Magic_Identity} it is shown that the non-local magic for a state close to a product state is related to the largest Schmidt weight through,
\begin{align}
    M_{\text{NL}}(|\text{out}\rangle)\ = \ 4\eta \ + \ O(\eta^2) \ . 
    \label{eq:NL_Magic}
\end{align}
The linear entropy used in the entanglement power is,
\begin{align}
   1 \ - \ \Tr\rho_A^2 \ = \ 1 \ - \ \sum_i \lambda_i^2 \ = \ 2\eta \ + \ O(\eta^2) \ ,
\end{align}
and therefore, to leading order, the non-local magic is twice the linear entropy.
This immediately implies that the Haar averaged non-local magic power is,\footnote{To this order, the same result holds for the product stabilizer ensemble used in Ref.~\cite{Robin:2026lnm} when $d$ is prime. 
For prime $d$ the single-qudit stabilizer states form a 2-design~\cite{WOOTTERS1989363,Klappenecker:2005gax}, and therefore the linear entropy averaged over independent product stabilizers equals the Haar average.}
\begin{align}
    {\cal P}_{\text{NL}}(S) \ = \ 2 {\cal E}_{\rm p}(S) \ = \ \frac{4\tau^2}{(d+1)^2}\Tr{\widetilde K^2} \ + \ O(\tau^4) \ .
\end{align}
Despite being distinct measures of quantum complexity, the entanglement and non-local magic power are simply related by a factor of two near the identity and SWAP.
In the next section, this proportionality is proven to hold for a broad class of non-local quantum resources.
The dependence on the phase shifts through $\Tr{\widetilde{K}^2}$ is shown to measure the squared projective Hilbert-Schmidt distance of the S-matrix from the closest non-entangling gate.

\section{Hilbert-Schmidt distance and the local universality of non-local resources}
\label{sec:HS_Metric}
The Hilbert-Schmidt metric gives the infinitesimal distance between two S-matrices as~\cite{Beane:2020wjl},
\begin{align}
    ds_{\text{HS}}^2 \ = \  \Tr \left [(S(\bm\delta) -S(\bm\delta'))(S(\bm\delta) -S(\bm\delta'))^{\dagger}\right ]  \ = \  4\sum_\mu r_\mu\,d\delta_\mu^2
    \label{eq:HS1}
\end{align}
where ${\bm \delta}'={\bm \delta}+d{\bm \delta}$.
Two S-matrices have equivalent action on Hilbert space if they only differ by an overall phase, i.e. $S({\bm \delta}')|\psi\rangle = e^{i\theta}S({\bm \delta})|\psi\rangle$.
Because of this, both the entanglement power and non-local magic are invariant under changes in the overall phase. 
The distance metric can be made to share this property by modifying Eq.~\eqref{eq:HS1} so that an overall phase does not change the distance.
This is achieved by making the overall phase a null direction
and pulling back the Hilbert--Schmidt metric to the space of relative phases,
\begin{equation}
  ds^2
  \ = \ 4\sum_\mu r_\mu\,d\delta_\mu^2
    \ - \ \frac4{d^2}\left(\sum_\mu r_\mu\,d\delta_\mu\right)^2 \ = \ \sum_{\mu\nu}C_{\mu\nu} d\delta_\mu d\delta_\nu \ .
  \label{eq:twometrics}
\end{equation}
The metric in phase coordinates has been defined as $C_{\mu \nu}$.

\begin{figure}[t]
 \centering
 \includegraphics[width=0.99\textwidth]{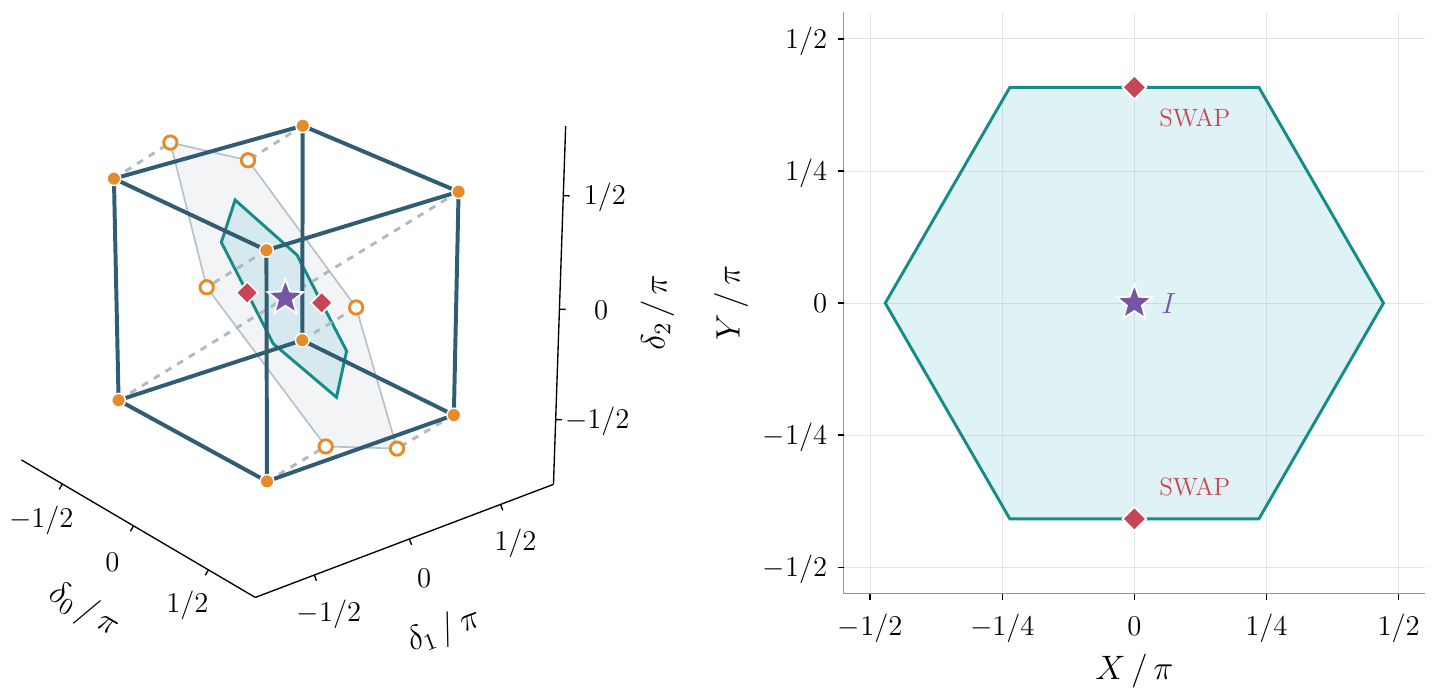}
 \caption{The space of S-matrices with $J=1$.
 Left: a full phase-shift cell,
$\delta_\mu\in[-\pi/2,\pi/2]$, with opposite faces identified.
Projection along the $(1,1,1)$ common-phase direction onto
$\delta_0+\delta_1+\delta_2=0$ produces the larger gray hexagon.
Under this identification, the filled orange circles are projected onto the unfilled circles.  
The blue region is one Voronoi cell of the projected
period lattice, with opposite edges identified.
The purple star represents the identity and the two red
diamonds represent SWAP.
Right: the same blue Voronoi cell in coordinates
$X=(\delta_2-\delta_0)/\sqrt{2}$ and
$Y=(2\delta_1-\delta_0-\delta_2)/\sqrt{6}$.}
 \label{fig:latticequotient}
\end{figure}

This projective geometry is illustrated for $J=1$ in
Fig.~\ref{fig:latticequotient}.
The space of S-matrices is parameterized
by three phase shifts $(\delta_0,\delta_1,\delta_2)$.
Each phase shift has period $\pi$ and this space can be
represented by a cube with opposite faces identified.
Identifying S-matrices that differ by an overall phase
removes the $(1,1,1)$ direction which is represented geometrically by projecting onto the plane
$\delta_0+\delta_1+\delta_2=0$.
The (Voronoi) unit cell of this projective geometry is denoted by the blue hexagon whose opposite edges are identified.

To make connection with Section~\ref{eq:Identity_Expand}, parameterize a small perturbation as $S({\bm \delta}') = S({\bm \delta)} e^{i\tau K}$.
The infinitesimal distance between $S({\bm \delta}')$ and $S({\bm \delta)}$ is
\begin{align}
    \ell^2
\ = \ \tau^2\Tr\widetilde K^2
\ .
\label{eq:HS_dist}
\end{align}
Applying this to $S({\bm \delta)}=S_0=I,\text{SWAP}$ reveals that the entanglement and non-local magic power can be expressed as
\begin{align}
 {\cal P}_{\text{NL}}(S) \ = \ 2 {\cal E}_{\rm p}(S) \ = \ \frac{4}{(d+1)^2}   \ell^2 \ + \ O(\ell^4) \ .
\end{align}
We emphasize that this leading $\ell^2$ response does not follow from a simple Taylor expansion of a resource function $R(S)$ around one of its zeros. 
This would generically have the form 
\begin{align}
    R\left (S_0e^{i\tau \widetilde{K}}\right ) \ = \ \frac{1}{2}\tau^2\sum_{\mu\nu}R_{\mu \nu}(S_0)\, \epsilon_\mu \epsilon_{\nu} \ + \ o(\tau^2) \ ,
\end{align}
where $R_{\mu\nu}(S_0)$ is the resource Hessian around $S_0$.
The entanglement and non-local magic power Hessian matrix is proportional to the Hilbert-Schmidt metric near the identity and SWAP, i.e. $R_{\mu\nu}(S_0) \propto C_{\mu \nu}$.
We prove this proportionality holds for any reasonably defined smooth non-local quantum resource in the following theorem:
\begin{theorem}
\label{theorem:Resource}
Consider a non-negative resource function $R[U(d^2)]\to \mathbb{R}_{\geq 0}$ with the following properties:
\begin{itemize}
    \item The resource is twice differentiable near the identity where it vanishes, $R(I)=0$.
    \item The resource is invariant under SWAP, global phases and local unitaries: \[R(U)=R(\text{SWAP}\,U)=R(e^{i\theta}U)=R[(u\otimes v)U(u^{\dagger}\otimes v^{\dagger})] \ . \]
\end{itemize}
The resource of a rotationally invariant S-matrix $S=S_0e^{i\tau \widetilde{K}}$ near $S_0=I$ or $S_0=SWAP$ is,
\begin{align}
    R(S) \ = \ \frac{1}{2} \chi_R \ell^2 \ + \ o(\ell^2)
    \label{eq:resource_HS}
\end{align}
where $\chi_R$ is a resource dependent proportionality constant and $\ell$ is the projective Hilbert-Schmidt distance from $S_0$ defined in Eq.~\eqref{eq:HS_dist}.
\end{theorem}
\noindent
The proof is given in Appendix~\ref{app:resource_proof}. The theorem exhibits a clear analogy with
Hooke's law, where $\ell$ is the displacement, $\chi_R$ is a kind of
stiffness, and $R$ is the stored energy.  All of the geometry 
resides in $\ell$ and is therefore shared among all
resources by construction.  Hence all of the resource dependence
resides in one number, $\chi_R$.

\begin{figure}[t]
 \centering
 \includegraphics[width=0.99\textwidth]{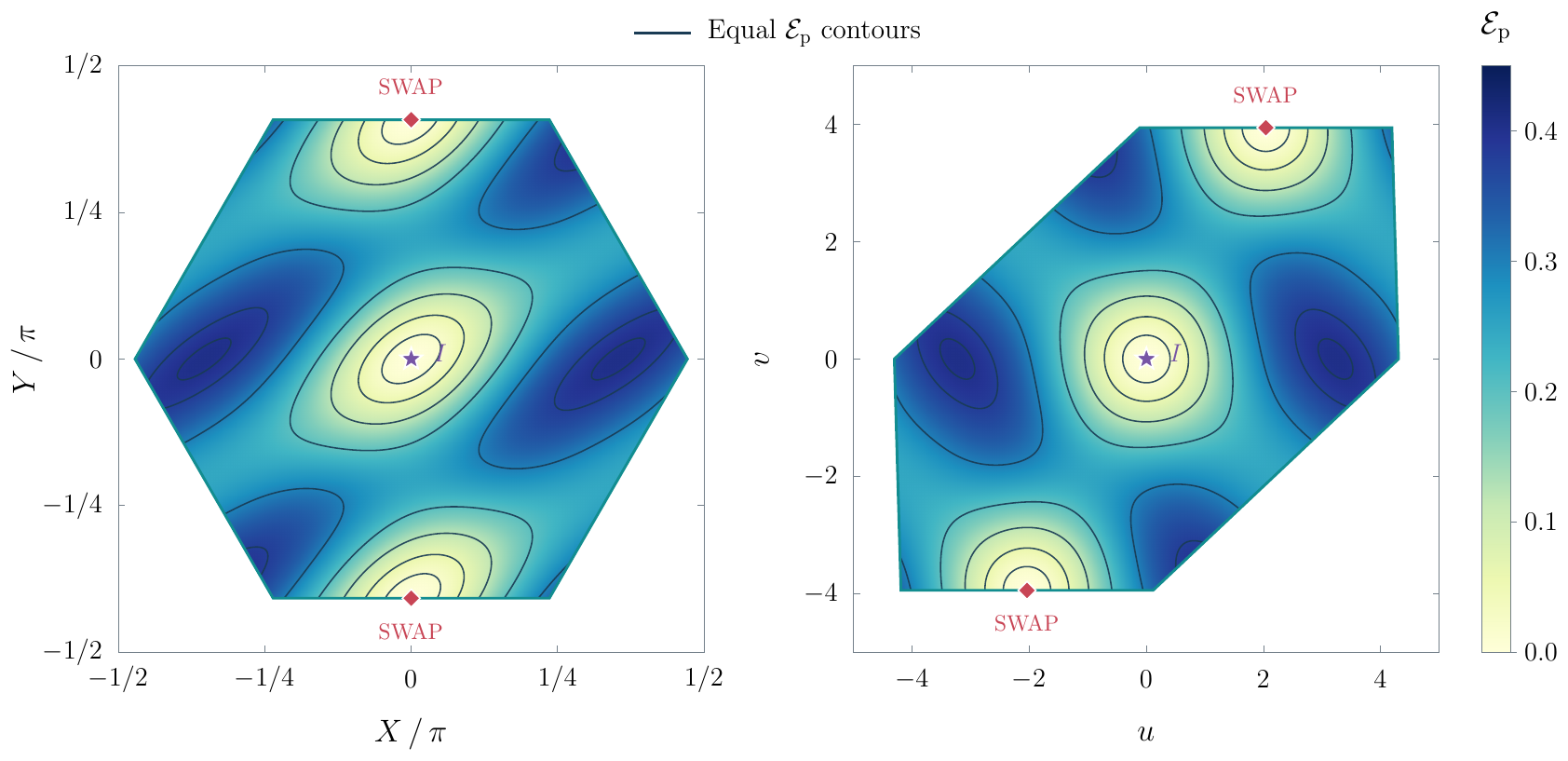}
 \caption{The entanglement power for $J=1$ S-matrices.
 Left: The entanglement power in the same Voronoi cell coordinates $(X,Y)$ used in Fig.~\ref{fig:latticequotient}.
Right: The entanglement power in coordinates $(u,v)$ that are orthonormal with respect to the Hilbert-Schmidt metric.}
 \label{fig:latticeEP}
\end{figure}
The proportionality of a quantum resource to the Hilbert-Schmidt distance only holds locally, and breaks down beyond $O(\ell^2)$.
This is illustrated in Fig.~\ref{fig:latticeEP} which plots the $J=1$ entanglement power throughout the Voronoi unit cell.
The $J=1$ entanglement power in relative coordinates $x=2(\delta_1 - \delta_0)$ and $y=2(\delta_2 - \delta_0)$ is~\cite{Beane:2021zvo},
\begin{align}
 {\cal E}_{\rm p}(S) \ = \ \frac1{648}\big[ &65(1-\cos y)
 +10(1-\cos2y)+60[1-\cos(2x-2y)]\nonumber \\
 &+15[1-\cos(2x-y)]+6(1-\cos2x)\big ] \ .
 \label{eq:spin1EP}
\end{align}
The left plot in Fig.~\ref{fig:latticeEP} uses the $(X,Y)$ coordinates in Fig.~\ref{fig:latticequotient} which are not orthonormal with respect to the metric in Eq.~\eqref{eq:twometrics}.
This obscures the connection between distance and entanglement.
In contrast, the right plot uses coordinates $(u,v)$ that are orthonormal with respect to the metric and the equi-entanglement contours (black lines) are constant-radii circles near identity and SWAP.

\section{Discussion}
\label{sec:discussion}
A guiding principle of effective field theory is that any operator allowed by symmetries will be present in the effective action.
It is often assumed that coefficients are of ``natural" size, i.e. $O(1)$ in the appropriate units.
In several cases, experimental data implies unnaturally small or large coefficients, with a prominent example being the large scattering lengths in
nucleon-nucleon scattering~\cite{Weinberg:1990rz,Weinberg:1991um,Kaplan:1996xu,vanKolck:1998bw,Kaplan:1998tg,Kaplan:1998we}.
These unnatural coefficients can lead to approximate symmetries that
are not exact symmetries of the fundamental theory, such as enhanced
spin-flavor symmetry~\cite{Mehen:1999qs,NPLQCD:2020lxg,Wagman:2017tmp} or non-relativistic conformal invariance~\cite{Mehen:1999nd,Nishida:2007pj}. 
Some approximate symmetries have been found to minimize the entanglement produced in scattering~\cite{Beane:2018oxh,Low:2021ufv,Beane:2021zvo,Liu:2022grf,Carena:2023vjc,Hu:2025lua,Kanemura:2026won}, leading to the conjecture that symmetry emerges from entanglement minimization.
The present work shows that there is nothing special about entanglement minimization, at least for rotationally invariant scattering.
Near the non-entangling points many non-local resources are proportional and suppressing entanglement suppresses every such resource by the same factor.
This applies for all $J$, and even to the non-local magic whose definition includes an optimization over local unitary transformations that can only be performed exactly for $J=1/2$~\cite{Qian:2025oit}.

This work demonstrates the value of an ``operator-centric" view of quantum resources, as opposed to a ``state-centric" view. 
Previous studies of the entanglement power in scattering computed the linear entropy for every possible initial state, and then integrated the general expression over a parameterized $\mathbb{CP}^{d-1}\times\mathbb{CP}^{d-1}$ manifold.
The complexity quickly grows with $J$ as is illustrated by the $J=3/2$ entanglement power derived in Ref.~\cite{Hu:2025lua} which is the sum of 14 trigonometric terms.
Rather than integrating over the non-linear state manifolds for each $J$, we specialize the operator formula from Ref.~\cite{PhysRevA.63.040304} using only $SU(2)$ representation theory.
The operator-centric perspective also provided insight into the universal form of non-local quantum resources near identity and SWAP.
We showed that the leading dependence is simply a measure of distance from the nearest resource-free gate in the space of all S-matrices.

\section*{Acknowledgments}

\noindent This work was supported by the U.~S.~Department of Energy
award {\bf DE-SC0020970} via the program on Quantum Horizons: QIS
Research and Innovation for Nuclear Science. R.F.~acknowledges support from the U.S.~Department of Energy QuantISED program through the theory consortium “Intersections of QIS and Theoretical Particle Physics” at Fermilab, from the U.S.~Department of Energy, Office of Science, Accelerated Research in Quantum Computing, Quantum Utility through Advanced Computational Quantum Algorithms (QUACQ), and from the Institute for Quantum Information and Matter, an NSF Physics Frontiers Center
(PHY-2317110). 
R.F.~additionally acknowledges support from a Burke Institute prize fellowship.
We acknowledge the use of OpenAI's GPT-6 Astra and Anthropic's Claude Opus 5.5 in the preparation of this manuscript.

\clearpage

\appendix

\section{Entanglement power for arbitrary spin}
\label{app:EP_SVD}
The entanglement power for a general unitary was derived in Refs.~\cite{PhysRevA.63.040304,Zanardi:2000zz} and is given in Eq.~\eqref{eq:EP_general}.
Applying this formula to the S-matrix requires extracting the singular values of the re-aligned coefficient matrix, ${\cal R}(S)$ which is obtained by regrouping the matrix indices.

The S-matrix is defined in the total spin basis in Eq.~\eqref{eq:Smatgate}.
A complete single particle basis is given by the irreducible spherical tensors $T_{LM}$ normalized such that,
\begin{align}
    \Tr{\left ( T_{LM}^{\dagger}T_{L'M'} \right )} \ = \ \delta_{LL'}\delta_{MM'} \ .
\end{align}
Specifically, one could write,
\begin{align}
    S \ = \ \sum_{L,L'=0}^{2J}\sum_{M=-L}^L\sum_{M'=-L'}^{L'} c_{LL'MM'} T_{LM}\otimes T_{L'M'} \ .
    \label{eq:SmatGeneral}
\end{align}
The allowed terms in the decomposition are strongly constrained by rotational symmetry.
Each total spin projector $P_{\mu}$ is rotationally invariant and can therefore be decomposed as,
\begin{align}
    P_\mu  \ = \ \sum_{L=0}^{2J} c_{L\mu}Q_L
\end{align}
where the $Q_L$ is the rotationally invariant product of the $T_{LM}$,
\begin{align}
  Q_L \ = \ \sum_{M=-L}^L T_{LM}\otimes T^{\dagger}_{LM} \ .  
\end{align}
The $Q_L$ are hermitian which can be shown using the angular momentum identity $T_{LM}^{\dagger} = (-1)^MT_{L-M}$.
The rotational invariance of the S-matrix therefore restricts the general form in Eq.~\ref{eq:SmatGeneral} to,
\begin{align}
    S \ = \ \sum_{\mu=0}^{2J} e^{2i \delta_{\mu}}\sum_{L=0}^{2J} c_{L\mu}Q_L \ \equiv \ \sum_{L=0}^{2J} a_L Q_L
    \label{eq:SmatQ}
\end{align}
where the second equality has exposed the singular values $a_L$ that are needed for the entanglement power.

Orthogonality of the $Q_L$ implies that the coefficients can be extracted as,
\begin{align}
    c_{L\mu} \ = \ \Tr \left (Q_L^{\dagger}P_\mu \right ) / \Tr\left (Q_L^{\dagger} Q_L\right ) \ = \ \frac{1}{2L+1}\Tr \left (Q_L^{\dagger}P_\mu \right ) \ ,
    \label{eq:Orthogonality}
\end{align}
where the $1/(2L+1)$ prefactor follows from the normalization of the $T_{LM}$.
The overlap is,
\begin{align}
    &\Tr \left (Q_L^{\dagger}P_\mu \right ) \ = \ \sum_{N=-\mu}^{\mu}\sum_{M=-L}^L  \langle\mu,N|T_{LM}\otimes T_{LM}^{\dagger} |\mu,N\rangle \nonumber \\
    &=  \sum_{N=-\mu}^{\mu}\sum_{M=-L}^L\sum_{m_1,m_2,m'_1,m'_2=-J}^J  \langle\mu,N|m_1;m_2\rangle \langle m_1;m_2|T_{LM}\otimes T_{LM}^{\dagger} |m'_1;m'_2 \rangle \langle m'_1;m'_2|\mu,N\rangle \nonumber \\
    &=\sum_{N=-\mu}^{\mu}\sum_{M=-L}^L\sum_{m_1,m_2,m'_1,m'_2=-J}^J  \langle\mu,N|m_1;m_2\rangle \langle m_1|T_{LM}|m'_1\rangle \langle m'_2|T_{LM} |m_2 \rangle \langle m'_1;m'_2|\mu,N\rangle \nonumber \\
    &= (2\mu+1)(2L+1)\sum_{N=-\mu}^{\mu}\sum_{M=-L}^{L}\sum_{m_1,m_2,m_1',m_2'=-J}^{J}
(-1)^{2J-m_1-m_2'}
\begin{pmatrix} J & J & \mu \\ m_1 & m_2 & -N \end{pmatrix}
\begin{pmatrix} J & J & \mu \\ m_1' & m_2' & -N \end{pmatrix}
\nonumber\\
&\qquad\times
\begin{pmatrix} J & L & J \\ -m_1 & M & m_1' \end{pmatrix}
\begin{pmatrix} J & L & J \\ -m_2' & M & m_2 \end{pmatrix} \nonumber \\
&= (2L+1)(2\mu+1)\,(-1)^{2J-\mu}
\begin{Bmatrix} J & J & \mu \\ J & J & L \end{Bmatrix} \ .
\label{eq:6j_math}
\end{align}
The first line has simplified the trace using the fact that $P_\mu$ is a projector and therefore only states with total spin $\mu$, $|\mu,N\rangle$ need to be summed.
The second line has inserted two complete sets of states in the single particle basis $|m,m'\rangle \equiv|J,m\rangle\otimes|J,m'\rangle$.
The fourth line has replaced the expectation values with the Wigner $3j$ symbol.
The fifth line has used the identity relating the Wigner $6j$ symbol in terms of products of Wigner $3j$ symbols given in section 34.4 of Ref.~\cite{NIST:DLMF}.
Identities related to permutation symmetries of the Wigner symbols have been used to put it into this form.
Inserting Eq.~\eqref{eq:6j_math} into Eq.~\eqref{eq:Orthogonality} and Eq.~\eqref{eq:SmatQ} gives the singular values as,
\begin{align}
    a_L \ = \ \sum_{\mu =0}^{2J}e^{2i\delta_{\mu}}(2\mu+1)(-1)^{2J-\mu}\begin{Bmatrix} J & J & \mu \\ J & J & L \end{Bmatrix} \ . 
    \label{eq:aL}
\end{align}

The other component of the entanglement power is $b_L$, the singular values obtained from $S\,\text{SWAP}$.
The SWAP gate is also diagonal in total spin, giving $(+1)$ for even-$2J-\mu$ and $(-1)$ for odd-$2J-\mu$.
Therefore,
\begin{align}
    S\,\text{SWAP} \ = \ \sum_{\mu=0}^{2J}(-1)^{2J-\mu} e^{2i\delta_{\mu}}P_{\mu}
    \label{eq:SSwap_decomp}
\end{align}
The singular value $b_L$ is obtained by replacing $e^{2i\delta_{\mu}}\to e^{2i\delta_{\mu}} (-1)^{2J-\mu}$ in Eq.~\eqref{eq:aL} giving,
\begin{align}
    b_L \ = \ \sum_{\mu =0}^{2J}e^{2i\delta_{\mu}}(2\mu+1)\begin{Bmatrix} J & J & \mu \\ J & J & L \end{Bmatrix} \ .
\end{align}
The entanglement power is thus Eq.~\eqref{eq:EP_anyJ} with $a_L$ and $b_L$ defined above.

The Wigner $6j$ symbol is given explicitly by the Racah formula, see section 34.4 of Ref.~\cite{NIST:DLMF},
\begin{equation}
\!\!\!\!\!
\begin{Bmatrix} J & J & \mu \\ J & J & L \end{Bmatrix} 
= f_L\, f_\mu
\sum_{t = 2J + \max(L,\mu)}^{\min(4J,\; 2J+L+\mu)}
\frac{(-1)^t\,(t+1)!}
     {\big[(t-2J-L)!\big]^2\,\big[(t-2J-\mu)!\big]^2\,(4J-t)!\,\big[(2J+L+\mu-t)!\big]^2} ,
\label{eq:racah-special}
\end{equation}
with
\begin{equation}
f_k \ = \  \frac{(2J-k)!\,(k!)^2}{(d+k)!} \ , \qquad 0 \le k \le 2J \ .
\end{equation}

\section{The entanglement and non-local magic power around the identity}
\label{app:EP_Magic_Identity}
For small perturbations from the non-entangling points, the S-matrix can be parameterized as $S=S_0 e^{i\tau \widetilde K}$ where $S_0=I,\text{SWAP}$.
Perturbations that simply change the overall phase of the S-matrix are not physical and we take $\Tr{\widetilde K}=0$.
The entanglement power contains ${\cal C}(S)+{\cal C}(S\,\text{SWAP})$ which is equal at both $S_0=I$ and $S_0=\text{SWAP}$ due to $(\text{SWAP})^2=I$. Therefore, without loss of generality, we restrict the following to $S_0=I$.

\subsection{Entanglement power}
From Eq.~\eqref{eq:CsCsSwap} and Eq.~\eqref{eq:SmatQ},
\begin{align}
    {\cal C}(S) \ = \  \sum_{L=0}^{2J}(2L+1)|a_L|^4 \ , \qquad S \ = \ \sum_{L=0}^{2J}a_LQ_L \ .
\end{align}
Applying the orthogonality condition, $\Tr{(Q_L Q_{L'}^{\dagger})}=\delta_{L,L'}(2L+1)$, gives,
\begin{align}
    a_L \ = \ \Tr{\left ( Q_L^{\dagger} S\right )}/(2L+1) \ = \ \Tr{\left [ Q_L^{\dagger} \left (I +i\tau \widetilde{K} - \frac{\tau^2}{2}\widetilde{K}^2 + O(\tau^3)\right  )\right ]}/(2L+1) \ .
\end{align}
The spin-0 scalar is $Q_0 = I/d$ with the overall normalization fixed by the orthogonality conditions.
Therefore,
\begin{align}
  |a_0|^4 \ = \ d^4 \ - \ 2d^2\tau^2 \Tr{\widetilde K^2}  \ + \ O(\tau^4) \ .  
\end{align}
Orthogonality to $Q_0$ implies that $\Tr{Q_{L\geq 1}} = 0$ and therefore the leading term for $L\geq 1$ is,
\begin{align}
    |a_{L\geq 1}|^4 \ = \ |i\tau \Tr{\left ( Q^{\dagger}_L \widetilde{K} \right )}/(2L+1)|^4 \ = \ O(\tau^4) \ .
\end{align}
Thus the leading order correction is entirely driven by $a_0$ and,
\begin{align}
    {\cal C}(S) \ = \  \left ( d^4 \ - \ 2d^2\tau^2 \Tr{\widetilde K^2} \right ) \ + \ O(\tau^4)
    \label{eq:CSd4}
\end{align}

Next is the evaluation of,
\begin{align}
    {\cal C}(S\,\text{SWAP}) \ = \  \sum_{L=0}^{2J}(2L+1)|b_L|^4 \ , \quad S\,\text{SWAP} \ = \ \sum_{L=0}^{2J}b_LQ_L \ .
    \label{eq:CSSWAP}
\end{align}
The coefficient is,
\begin{align}
    b_L \ = \ \Tr{\left [Q_L^{\dagger}\,\text{SWAP} \left (I+i\tau \widetilde{K} -\frac{\tau^2}{2}\widetilde{K}^2+O(\tau^3) \right )\right ]} /(2L+1) \ .
    \label{eq:bLExpansion}
\end{align}
We will show that only the identity term in the expansion contributes at this order.
A useful fact is that the SWAP operator can be expressed as the sum over all tensor product operators in an orthonormal basis~\cite{Nambu:2005amn}.
This implies that
\begin{align}
    \text{SWAP} \ = \ \sum_{L=0}^{2J} Q_L \ ,
\end{align}
and therefore the leading term is
\begin{align}
     \Tr{\left (Q_L^{\dagger}\,\text{SWAP}  \right )}/(2L+1) \ = \ 1 \ .
\end{align}
The next-to-leading term is,
\begin{align}
     i\tau\Tr{\left (Q_L^{\dagger}\,\text{SWAP}\, \widetilde{K} \right )}/(2L+1) \ \equiv \ i \tau \beta_L \ .
\end{align}
This contribution is purely imaginary because the matrices inside the trace are hermitian and commuting.
Finally the next-to-next-to-leading term is
\begin{equation}
  -\frac{\tau^2}{2}
  \Tr
  \left(
    Q_L^\dagger\,\text{SWAP}\,\widetilde{K}^2
  \right)/(2L+1) \ \equiv \  -\frac{\tau^2}{2} \gamma_L \ , 
\end{equation}
and therefore
\begin{equation}
   b_L
  \ = \ 
  1\ + i\tau\beta_L-\frac{\tau^2}{2}\gamma_L
  \ + \ O(\tau^3) \ .
\end{equation}

Inserting this into Eq.~\eqref{eq:CSSWAP} gives
\begin{align}
   {\cal C}(S\,\text{SWAP}) \ 
   &= \ \sum_{L=0}^{2J}(2L+1)\left [ 1 + 2\tau^2 (\beta_L^2-\gamma_L) + O(\tau^4)\right ] \nonumber \\
   &= \ d^2 \ + \ 2\sum_{L=0}^{2J}(2L+1)\tau^2 (\beta_L^2-\gamma_L) \ + \  O(\tau^4) \ .
\end{align}
The $O(\tau^2)$ term is exactly zero due to the normalization condition,
\begin{align}
    \underbrace{\Tr{\left ((S\,\text{SWAP})(S\,\text{SWAP})^{\dagger} \right )}}_{=d^2}  \ &= \ \Tr{\left [ \left (\sum_{L=0}^{2J}b_LQ_L\right )\left (\sum_{L'=0}^{2J}b_{L'}^*Q^{\dagger}_{L'}\right ) \right ]}
 \nonumber \\
 &= \ \sum_{L=0}^{2J}(2L+1)|b_L|^2 \nonumber \\
 & = \ \sum_{L=0}^{2J}(2L+1)\left [1+\tau^2(\beta_L^2 -\gamma_L)\right ] \ + \  O(\tau^4) \nonumber \\
 & = \ d^2 + \underbrace{\sum_{L=0}^{2J}(2L+1)\tau^2(\beta_L^2-\gamma_L)}_{=0} \ + \  O(\tau^4)  \ . 
\end{align}
It therefore follows that
\begin{align}
    {\cal C}(S\,\text{SWAP}) \ = \  d^2 \ + \  O(\tau^4) \ .
    \label{eq:CSSWAP_d2}
\end{align}
Combining Eq.~\eqref{eq:CSSWAP_d2} and Eq.~\eqref{eq:CSd4} gives the entanglement power in Eq.~\eqref{eq:EP_LO}.

\subsection{Non-local magic power}
The goal of this section is to show the relation between the non-local magic and the largest Schmidt weight in Eq.~\eqref{eq:NL_Magic}.
This proceeds in the following steps:
\begin{enumerate}
    \item[1.] The linear magic of a state near a stabilizer state $|s\rangle$ is $M_{\text{lin}}\left ( \sqrt{1-\epsilon}|s\rangle+\sqrt{\epsilon}|w\rangle\right ) = 4\epsilon +O(\epsilon^2)$.
    \item[2.] For an S-matrix near the identity, $|\text{out}\rangle$ is close to a product state.
    The local unitaries that minimize the linear magic bring the state close to a product stabilizer. 
    \item[3.] The smallest achievable weight outside a product stabilizer is $1 - \max_{x,y}|\langle x|\langle y|\text{out}\rangle|^2 = 1 - \lambda_0$ where $\lambda_0$ is the largest Schmidt weight.
    \item[4.] Therefore, to lowest order, $M_{\text{lin}}\left ( |\text{out}\rangle\right ) = 4(1-\lambda_0)$.
\end{enumerate}
Before establishing [1.] we state our conventions.
The representative elements of the Pauli-Weyl group are chosen to be,
\begin{align}
    \mathcal{P}_D= \{X^{a}Z^{b}\otimes X^{c}Z^{e} : a,b,c,e\in\{0,1,\dots,d-1\}\} \ , \quad X|j\rangle = |j+1\text{ mod }d\rangle\ , \quad Z|j\rangle = e^{2\pi i j/d}|j\rangle \ .
\end{align}
There are $D^2$ Pauli operators modulo scalar phases, and a  stabilizer state $|s\rangle$ is an eigenstate of exactly $D$ of them. 
Choosing their phases so that each eigenvalue is $(+1)$ gives the stabilizer group ${\mathcal S}_s$. The remaining $D^2-D$ Pauli operators map $|s\rangle$ to states orthogonal to $|s\rangle$, and therefore have vanishing expectation values.
Below we will use the fact that the stabilizer subgroup forms a projector onto the stabilizer state,
\begin{align}
    \frac{1}{D}\sum_{P\in {\cal S}_s} P\ = \ |s\rangle\langle s| \ .
    \label{eq:stab_projector}
\end{align}

To show [1.], consider a state close to a stabilizer,
\begin{align}
    |\Phi\rangle \ = \ \sqrt{1-\epsilon}|s\rangle \ + \ \sqrt{\epsilon }|w\rangle \ , \quad \langle w|s\rangle \ = \ 0 \ .
    \label{eq:almost_stab}
\end{align}
The linear magic is,
\begin{align}
    M_{\text{lin}}(|\Phi\rangle) \ &= \ 1 - \ \frac{1}{D}\sum_{P\in {\cal P}_D} |\langle \Phi|P|\Phi\rangle|^4 \nonumber \\
    &= \ 1 - \ \frac{1}{D}\sum_{P\in {\cal S}_s}\left (1-4\epsilon+4\epsilon\text{Re}[\langle w |P|w\rangle]+O(\epsilon^2) \right ) \nonumber \\
    &= \ 4\epsilon\  + \ O(\epsilon^2) \ . 
\end{align}
The second line has used the defining property of a stabilizer state as well as $\langle s|w\rangle=0$.
The third line has used the projector identity in Eq.~\eqref{eq:stab_projector} along with $\langle s|w\rangle=0$.
This expansion was derived for qubit systems in Ref.~\cite{Huang:2026mhv}.

The outgoing state in its Schmidt basis is,
\begin{align}
|\mathrm{out}\rangle=\sum_i\sqrt{\lambda_i}|u_i\rangle|v_i\rangle \ ,
\qquad \lambda_0=1-\eta \ . 
\end{align}
Near the identity, the outgoing state is nearly a product state, so \(\eta\ll1\). We can choose local unitaries that rotate the dominant Schmidt product \(|u_0\rangle|v_0\rangle\) into a product stabilizer, such as \(|00\rangle\). The rotated state has the form
\begin{align}
    (U_A\otimes U_B)|\mathrm{out}\rangle
\ = \ \sqrt{1-\eta}\,|00\rangle+\sqrt{\eta}\,|w\rangle\ ,
\qquad \langle00|w\rangle=0 \ .
\end{align}
The stabilizer expansion derived above therefore gives $M_{\mathrm{lin}}=4\eta+O(\eta^2)$, providing an upper bound on the non-local magic.

Next, we argue that this choice of local rotation is optimal at lowest order in $\eta$. Since linear magic vanishes only on stabilizer states, any minimizing choice must place the outgoing state near a stabilizer. 
This stabilizer must be a product state as entangled stabilizers have a flat nonzero Schmidt spectrum and hence a largest Schmidt weight at most $1/2$, whereas local unitaries preserve the outgoing state's largest weight $1-\eta$.
Near a stabilizer $|s\rangle$, the leading magic is $4\epsilon$, where
\begin{align}
   \epsilon\ = \ 1-|\langle s|(U_A\otimes U_B)|\mathrm{out}\rangle|^2 \ . 
\end{align}
Thus, at leading order, minimizing magic amounts to maximizing the overlap with a product stabilizer. 
Local unitaries allow any product state to be rotated into such a stabilizer, and the product state that has the largest overlap with $|\text{out}\rangle$ is its largest Schmidt vector
\begin{align}
   \max_{|x\rangle,|y\rangle}
|\langle x|\langle y|\mathrm{out}\rangle|^2
\ = \ \lambda_0 \ = \ 1-\eta \ . 
\end{align}
This can be shown by considering the overlap with an arbitrary product state $|x\rangle|y\rangle$,
\begin{align}
    |\langle x|\langle y|\text{out}\rangle| \ &= \ \bigg |\sum_i \sqrt{\lambda_i}\langle x |u_i\rangle\langle y|v_i\rangle \bigg | \nonumber  \\
    &\leq \ \sqrt{\lambda_0}\sum_i |\langle x|u_i\rangle ||\langle y|v_i\rangle| \nonumber \\
     &\leq  \ \sqrt{\lambda_0} \left (\sum_i |\langle x|u_i\rangle |^2 \right )^{1/2}\left (\sum_i |\langle y|v_i\rangle |^2 \right )^{1/2}\nonumber \\
    &\leq  \ \sqrt{\lambda_0} \ .
\end{align}
The second line uses the triangle inequality and $\lambda_0 \geq \lambda_{i>0}$, the third line uses the Cauchy-Schwarz inequality and the fourth line uses orthonormality of the Schmidt basis.
This bound is saturated by $|x\rangle|y\rangle=|u_0\rangle|v_0\rangle$ and therefore the magic is minimized for $U_A\otimes U_B|u_0\rangle|v_0\rangle=|s\rangle$ where $|s\rangle$ is a product stabilizer.
Thus,
\begin{align}
  M_{\mathrm{NL}}(|\mathrm{out}\rangle)
\ = \ 4\eta\ + \ O(\eta^2) \ .  
\end{align}

\section{Local universality of non-local quantum resources}
\label{app:resource_proof}
This appendix provides the proof of Theorem~\ref{theorem:Resource}. Schur's lemma is used twice which is stated here:
\begin{lemma}[Schur's lemma]
\label{lem:schur}
Let a group $G$ with elements $g\in G$ act linearly on a finite-dimensional vector space $V$ through
operators $\rho(g)$, such that the only subspaces mapped into themselves by every
$\rho(g)$ are $\{0\}$ and $V$. 
Let $f:V\to V$ be a linear map that commutes with
the group action,
\begin{equation}
  f\big(\rho(g)\,v\big) \ = \  \rho(g)\,f(v)
  \qquad \text{for all } g\in G,\ v\in V \ .
  \label{eq:commutativity}
\end{equation}
If $V$ is complex, or if $V$ is real and $f$ is symmetric with respect to an inner
product on $V$, i.e.\ $\langle v, f(w)\rangle = \langle f(v), w\rangle$ for all $v,w\in V$, then there exists a constant $c$ such that
\begin{equation}
  f(v) \  =  \ c\,v \qquad \text{for all } v\in V \ . 
\end{equation}
\end{lemma}
\noindent
The proof below shows that a non-local quantum resource with the properties listed in Theorem~\ref{theorem:Resource} has the expansion in Eq.~\eqref{eq:resource_HS} for the region around the identity, $S=e^{i\tau K}$.
It immediately extends to the region around SWAP due to the assumed invariance of the resource function $R(S) = R(\text{SWAP}\,S)$.
The trace can be removed without loss of generality, $K\to\widetilde{K}$, because of the property $R(S) = R(Se^{i\theta})$.

Define $f(X) = R(e^{i X})$ for every generator $X$.
The twice-differentiable assumption around the identity as well as $f\geq 0 $ and $f(0)=0$ imply that,
\begin{align}
    f(X) \ = \ \frac{1}{2}B(X,X)  \ + \ o(||X||^2)
\end{align}
where $B$ is the Hessian of $f$ around zero.
The goal is to show that the Hessian is proportional to the squared Hilbert-Schmidt distance for $X = \tau \widetilde{K}$. 

Define $W = u\otimes v\in U(d)\otimes U(d)$.
The local invariance property is $R(W S W^{\dagger}) = R(S)$ which implies that this property is shared by $B$,
\begin{align}
    B(WXW^{\dagger}, WXW^{\dagger}) \ = \ B(X,X) \ .
\end{align}
This follows from $W e^{i X}W^{\dagger} = e^{i W X W^{\dagger}}$.
This invariance extends to 
\begin{align}
    B(WXW^{\dagger},WYW^{\dagger}) \ &= \ B(X,Y) \ .
\end{align}
To show this, note that bilinearity and symmetry give the identity
\[
B(X,Y)\ = \ \frac{1}{4}\big[B(X+Y,X+Y)-B(X-Y,X-Y)\big ]\ .
\]
Applying the linear map $X\to WXW^\dagger$ gives,
\begin{align}
    B(WXW^\dagger,WYW^\dagger)
 &= \tfrac14\big[B\big(W(X+Y)W^\dagger,W(X+Y)W^\dagger\big)-B\big(W(X-Y)W^\dagger,W(X-Y)W^\dagger\big)\big] \nonumber \\
&= \ \tfrac14\big[B(X+Y,X+Y)-B(X-Y,X-Y)\big]   \nonumber \\
& = \ B(X,Y) \ .
\label{eq:prop1}
\end{align}
This property will be used below.

Schur's lemma can be applied to the $SU(2)$ group of spin rotations acting on individual spins to show that $\Tr_B{\widetilde{K}} = cI$.
The traceless condition $\Tr{\widetilde{K}}=0$ implies $c=0$ giving the desired intermediate result $\Tr_B{\widetilde{K}} = 0$.
In this application of Schur's lemma,
$g=e^{i\vec{\theta}\cdot \vec{J}} \in G$ which acts on the state space of particle A, i.e. $V=\mathbb{C}^d$.
The linear map on $V$ is $f=\Tr_B{\widetilde{K}}$.
Particle A is in the spin-$J$ representation which is irreducible.
To show the required commutativity condition in Eq.~\eqref{eq:commutativity}, first note that rotational invariance implies that $S$, and therefore also $\widetilde{{K}}$, commute with total spin,
\[
[\widetilde{K},\vec{J}\otimes I + I\otimes \vec{J}] \ = \  0 \ . 
\]
Taking the partial trace of both sides gives,
\[
[\Tr_{B}\widetilde{K},\vec{J}] \ = \ 0 \ .
\]
Thus,
\begin{align}
   e^{i\vec{\theta}\cdot \vec{J}}\left ( \Tr_B{\widetilde{K}}\right ) v \ = \ \left ( \Tr_B{\widetilde{K}}\right )e^{i\vec{\theta}\cdot \vec{J}}v 
\end{align}
and Eq.~\eqref{eq:commutativity} is satisfied.
Schur's lemma gives,
\begin{align}
  \left ( \Tr_B{\widetilde{K}}\right )v \ = \ c \, v \quad \text{for all }v\in V \ ,  
\end{align}
and $\Tr_B{\widetilde{K}}\propto I$. The traceless condition fixes $c=0$ since,
\begin{align}
  \Tr_{A}\Tr_B \widetilde{K} \ = \ \Tr\widetilde{K}  \ = \ c\Tr{I} \ = \ cd \ = \  0 \ .  
\end{align}
The same holds if particle A and B are switched and
\begin{align}
    \Tr_{A}\widetilde{K} \ = \ \Tr_{B}\widetilde{K} \ = \ 0 \ .
    \label{eq:prop2}
\end{align}
We define the space of all such operators,
\begin{align}
  V_{\text{nl}} \ = \ \{X=X^{\dagger}: \Tr_{A}{X} =  \Tr_{B}{X} = 0\} \ .  
\end{align}

Building off Eq.~\eqref{eq:prop2}, we will now use Schur's lemma to prove that $B(X,X) \propto \Tr{X^2}$. Let $G=U(d)\times U(d)$ and $X\in V_{\text{nl}}$ such that $\rho(g) X \ = \ (u\otimes v) X (u^{\dagger}\otimes v^{\dagger})$.
The space $V_{\text{nl}}$ transforms in the irreducible $\text{adjoint}\otimes\text{adjoint}$ representation
of $U(d)\times U(d)$.
Without Eq.~\eqref{eq:prop2}, $X$ could contain components proportional to $I$, $A\otimes I$ or $I\otimes B$ which form separate invariant subspaces under $G$.
Schur's lemma would then allow for distinct proportionality constants in each subspace and the Hessian would not be proportional to the Hilbert-Schmidt distance.
Note that $\rho(g)$ maps $V_{\text{nl}}\to V_{\text{nl}}$ since it preserves the vanishing partial traces%
\begin{align}
    \Tr_B \left ( (u\otimes v) X (u^{\dagger}\otimes v^{\dagger}) \right ) \ = \ u\Tr_B\left ( (I\otimes v )X (I\otimes v^{\dagger})\right )u^{\dagger} \ = \ 0 \ .
\end{align}
It will be convenient to define an orthonormal basis $E_k\in V_{\text{nl}}$ with $\Tr(E_k E_l) = \delta_{kl}$.
Any element can be expanded as $X=\sum_k x_kE_k$ with real coefficients $x_k=\Tr(E_kX)$ and this space is isomorphic to $\mathbb{R}^{n}$ with $n=(d^2-1)^2$.
The group elements are $(d^2-1)^2\times(d^2-1)^2$ real matrices
$[\rho(g)]_{kl}=\Tr\big(E_k\,\rho(g)E_l\big)$ that act on vectors, $\rho(g)X=\sum_{kl}E_k[\rho(g)]_{kl}x_l$.

The linear map in Schur's lemma is the Hessian acting on $\mathbb{R}^{n}$.
Bilinearity of $B(X,Y)$ gives
\begin{align}
    B(X,Y) \ = \ \sum_{kl} x_k\,H_{kl}\,y_l \ ,
\qquad
H_{kl} \ = \ B(E_k,E_l) \ .
\end{align}
The Hessian matrix $H_{kl}$ is real and symmetric because $B$ is symmetric and the $X$ are hermitian. 
The goal is to show $H_{kl} \propto\delta_{kl}$ which would give $B(X,X)\propto\Tr(X^2)$. 
We next show the commutativity relation in Schur's lemma, $[H,\rho(G)]=0$, is satisfied.

In Eq.~\eqref{eq:prop1}, it was shown that the Hessian is invariant under local transformation,
\begin{align}
    B(X,Y) \ = \ B(\rho(g)X ,\rho(g)Y) \quad \text{for all }X,Y \ .
\end{align}
In index notation,
\begin{align}
    \sum_{kl}H_{kl}x_k y_l \ = \ \sum_{ijkl} [\rho(g)]_{i k} H_{ij} [\rho(g)]_{jl} x_k y_l \quad \text{for all }x_k,y_l \ .
\end{align}
Thus, as an $n$-dimensional matrix, 
\begin{align}
 H  \ = \  [\rho(g)]^TH[\rho(g)] \ .
 \label{eq:H_transform}
\end{align}
In this basis $\rho(g)$ preserves vector norms,
\begin{align}
    \sum_i\left (\sum_{j} [\rho(g)]_{ij}x_j\right )^2 \ = \ \Tr\big[(\rho(g)X)^2\big]
\ = \ \Tr\big[(u\otimes v)X^2(u\otimes v)^\dagger\big] \ = \ \Tr X^2 \ = \ \sum_k x_k^2 \ ,
\end{align}
so $[\rho(g)]$ is an orthogonal matrix, $[\rho(g)]^T=[\rho(g)]^{-1}$.
The condition in Eq.~\eqref{eq:H_transform} can be rewritten as $[\rho(g)]H = H [\rho(g)]$ and hence the commutator vanishes, $[H,[\rho(g)]]=0$.
Schur's lemma immediately implies that $H_{kl}\propto\delta_{kl}$ and,
\begin{align}
    B(X,Y) \ \propto \ \sum_{kl}x_k \delta_{kl}y_l \ . 
\end{align}
Denoting the proportionality constant as $\chi_R$ gives,
\begin{align}
    B(X,Y) \ = \ \chi_R \Tr{(XY)} \ .
\end{align}
Setting $X=\tau \widetilde{K}$ gives the desired form,
\begin{align}
    R\left (e^{i\tau \widetilde{K}}\right ) \ = \ \frac{\tau^2}{2}\chi_R \Tr{\widetilde{K}^2} \ + \ o(\tau^2) \ = \ \frac{1}{2}\chi_R \ell^2  \ + \ o(\ell^2) \ ,
\end{align}
where $\ell$ is the Hilbert-Schmidt distance from the identity derived in Section~\ref{sec:HS_Metric}.
\bibliographystyle{JHEP}
\bibliography{bibi}
\end{document}